\documentclass[preprint2]{aastex}
\usepackage{epsfig}
\usepackage{psfig}
\usepackage{natbib}

\newcommand{\snr}{G38.7--1.4}

\newcommand{\fcgs}{${\rm erg}\, {\rm cm}^{-2}\, {\rm s}^{-1}$}

\begin{document}

\title{Identification campaign of supernova remnant candidates in the Milky Way. II. X-ray studies of G38.7-1.4}
    
\author{R. H. H. Huang\altaffilmark{1},  J. H. K. Wu\altaffilmark{1}, C. Y. Hui\altaffilmark{2,4}, 
             K. A. Seo\altaffilmark{2}, L. Trepl\altaffilmark{3}, A. K. H. Kong\altaffilmark{1}}
\email{cyhui@cnu.ac.kr}
\altaffiltext{1}
{Institute of Astronomy and Department of Physics, National Tsing Hua University, Hsinchu, Taiwan}
\altaffiltext{2}
{Department of Astronomy and Space Science, Chungnam National University, Daejeon, Korea}
\altaffiltext{3}
{Astrophysikalisches Institut und Universit\"ats-Sternwarte, Universit\"at Jena, Jena, Germany}
\altaffiltext{4}
{Corresponding author}

\begin{abstract}
We report on XMM-Newton and Chandra observations of the Galactic supernova remnant candidate 
\snr, together with complementary radio, infrared, and $\gamma$-ray data. 
An approximately elliptical X-ray structure is found to be well correlated with radio shell as seen by 
the Very Large Array. The X-ray spectrum of \snr\ can be well-described by an absorbed collisional 
ionization equilibrium plasma model, which suggests the plasma is shock heated. 
Based on the morphology and the spectral behaviour, we suggest that \snr\ is indeed a supernova remnant belongs 
to a mix-morphology category. 
\end{abstract}


\section{Introduction}

Supernovae (SNe) and their remnants play a crucial role in driving the dynamical and chemical 
evolution of galaxies (Woosley \& Weaver 1995). 
Each SN produces bulks of heavy elements and disperses them throughout 
the interstellar medium (ISM) (e.g. Thielemann et al. 1996; Chieffi \& Limongi 2004). 
The shock waves from the SN explosions may also  
trigger star formation in molecular clouds (Boss 1995). In addition, the blast waves in supernova remnants (SNRs) 
can accelerate particles to relativistic energies via Fermi-I acceleration (Reynolds 2008), 
which has long been suggested as a promising acceleration mechanism for Galactic cosmic rays (GCRs). 
For investigating the role of SNRs as GCR accelerators, it is necessary to ask whether they can account
for the entire energy density of CRs in the Milky Way. This is related to the mechanical power provided
by the SNe, which in turn is associated with their event rate. In our Milky Way, the currently known 
SNR population is far below the expectation based on a event rate of 2 SNe/century (Tammann et al. 1994)
and a typical evolution timescale of $\sim10^{5}$~yrs (see Hui et al. 2012; Hui 2013). Therefore, deeper 
searches for Galactic SNRs are certainly needed. \\[-2ex]

With the much improved spatial and spectral resolution and enlarged effective area, 
state-of-the-art X-ray observatories like Chandra and XMM-Newton provide powerful 
tools for studying the shock-heated plasma and the non-thermal emission from the leptonic 
acceleration in SNRs (Hui 2013; Kang 2013). However, the number of X-ray detected SNRs 
is still significantly smaller than the corresponding number of detections in the radio. Until now 
there are 302 SNRs that have been uncovered in the Milky Way: 274 objects recorded in 
Green (2009) plus 28 objects reported in Ferrand \& Safi-Harb (2012), while the number of Galactic 
SNRs detected in X-rays is about 100\footnote{http://www.physics.umanitoba.ca/snr/SNRcat/}.
Therefore, enlarging the sample of the X-ray detected SNRs would be valuable. 
Recently, we have initiated an observational campaign for searching and identifying new Galactic SNRs 
with X-ray telescopes (Hui et al. 2012). Here we report our detailed X-ray analysis of the another 
SNR candidate \snr. \\[-2ex]

\snr\ was first detected as an unidentified object in ROSAT All-Sky Survey (RASS) with an extent of 
about $12' \times 8'$. This object has a centrally-peaked morphology in X-rays. Cross-correlating with the  
Effelsberg Galactic Plane 11~cm survey data, \snr\ is found to positionally coincide with an incomplete 
radio shell (cf. Figure~1d in Schaudel et al. 2002). A follow-up observation with the Effelsberg 
telescope at a wavelength of 6~cm revealed a non-thermal radio emission with a spectral index 
of $\alpha =-0.79\pm0.23$. Furthermore, the existence of polarization in the radio shell was 
reported by Schaudel et al. (2002). All these evidences suggest that \snr\ is very likely to be a 
SNR with center-filled X-ray (mixed) morphology. However, the poor spatial resolution ($\sim96$") 
and the limited photon statistic ($\sim50$ source counts) of RASS data do not allow one to 
unambiguously confirm its physical nature. 
Furthermore, the limited energy bandwidth of ROSAT (0.1-2.4~keV) does not allow one to determine 
whether a possible hard X-ray ($>$2~keV) component, arising from the interactions of the reflected 
shocks with the dense ambient medium or alternatively from the synchrotron emission radiated by 
relativistic leptons, is present in the hard X-ray band. This motivates us to carry out a detailed X-ray 
studies of \snr\ with XMM-Newton and Chandra. \\[-2ex]

Considering the composition of GCRs, leptons only constitute a small proportion. A large fraction of the 
observed GCRs are hadrons (i.e proton and heavy ions; cf. Sinnis et al. 2009). Due to the large masses of hadrons, they 
are not efficient synchrotron emitters. X-ray and radio observations are generally difficult to constrain 
their presence. On the other hand, the collision of relativistic hadrons can lead to the production of 
neutral pions which can subsequently decay into $\gamma-$rays (Caprioli 2011). For complementing the aforementioned 
X-ray investigation of \snr\ as a possible acceleration site of GCRs, we have also conducted a search for  
$\gamma-$ray emission at the location of \snr\ wtih the Large Area Telescope (LAT) on board the Fermi 
Gamma-Ray Space Telescope. \\[-2ex]

In this paper, we report a detailed high energy investigation of \snr. The observations and the data reduction of 
XMM-Newton and Chandra observatories are described in section 2. Sections 3 and 4 present the results of the X-ray 
spatial and spectral analysis respectively. In section 5, we describe a deep search of $\gamma-$ray emission with Fermi. 
Finally, we discuss the physical implications of the results and summarize our conclusions in sections 6 and 7 respectively.


\section{Observations \& Data Reduction}
\subsection{XMM-Newton Observation}

We have observed \snr\ with XMM-Newton on April 19, 2012 (ObsID: 0675070401) for a 
$\sim$20~ks total exposure. In this observation, the EPIC MOS1/2 and PN instruments were 
operated in full-frame mode using the medium filter to block optical stray light. 
All the data were processed with the XMM-Newton Science Analysis Software (SAS) package 
(Version 11.0.0). Calibrated event files for the MOS1, MOS2, and PN detectors were produced
using the SAS task {\em emchain} and {\em epchain}, following standard procedures. Events 
spread at most in two contiguous pixels for the PN (i.e., pattern = 0--4) and in four contiguous 
pixels for the MOS (i.e., pattern = 0--12) have been selected. 
We further cleaned the data by accepting only the good time intervals (GTIs) when the 
sky background was low for the whole camera ($<2.6$~counts~s$^{-1}$ for MOS1 and MOS2, 
$<20$~counts~s$^{-1}$ for PN). 
The effective exposure times after background cleaning for MOS1, MOS2, 
and PN are 11.0~ks, 12.3~ks, and 11.2~ks, respectively. Data analyses were restricted to the 
0.5--10.0~keV energy band.

\subsection{Chandra Observation}
We have also observed \snr\ with Chandra on 2012 June 9-10 (ObsID 13770)
using Advanced CCD Imaging Spectrometer (ACIS-I) with a frame time of 3.2~s. The total
exposure time is $\sim 28$~ks. The data was configured in the VFAINT telemetry mode. Data 
reduction and analysis were processed with Chandra Interactive Analysis Observations (CIAO) 
version 4.5 software and the Chandra Calibration Database (CALDB) version 4.5.5.1. 
The level-2 data with background cleaning was used in our study. Data analysis was restricted 
to the energy range of $0.5-8.0$\,keV.


\section{Spatial Analysis}

\begin{figure*}
\centerline{\psfig{figure=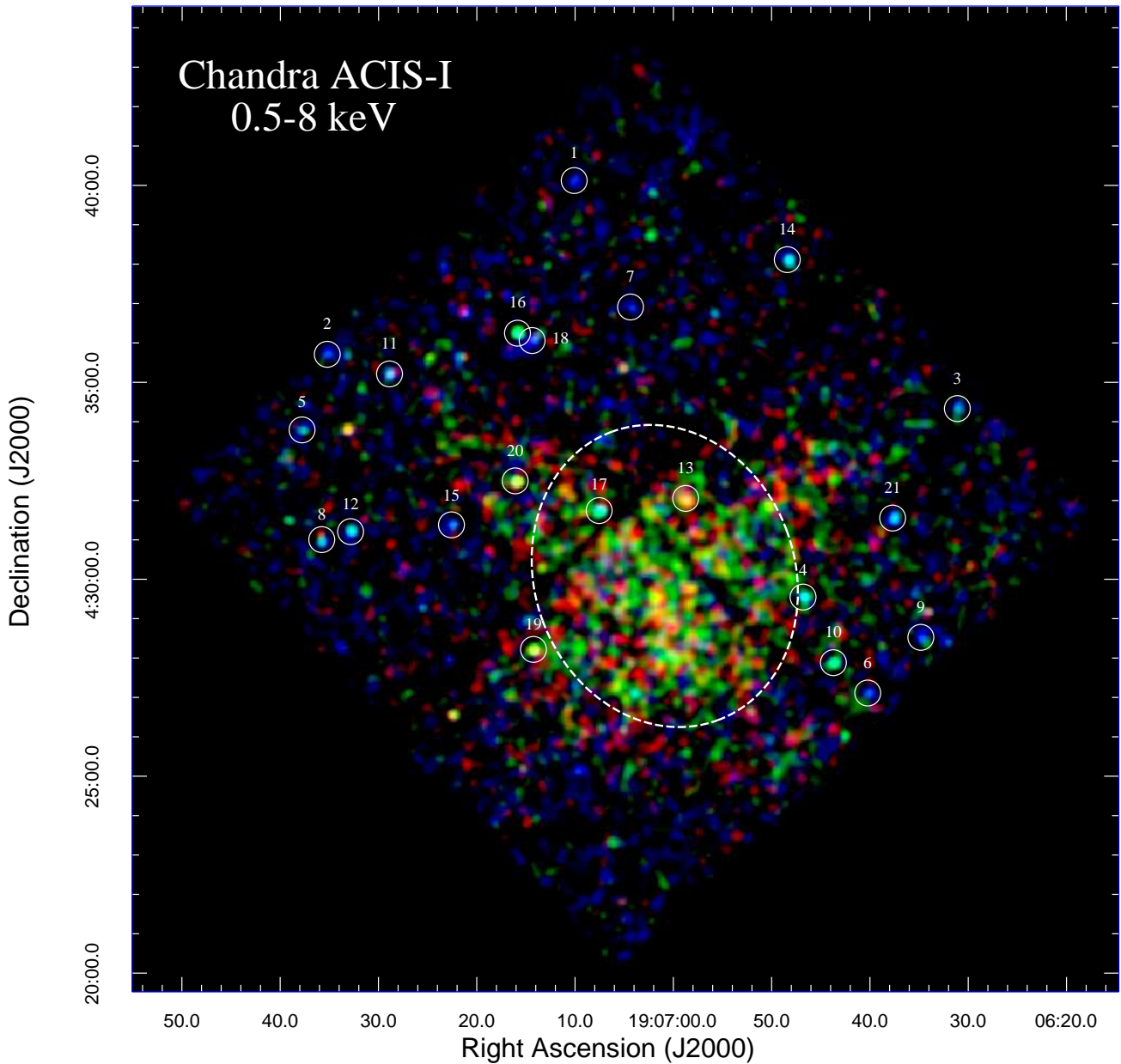,width=19cm,clip=}}
\caption{
A $25'\times25'$ X-ray color image of \snr\ as observed
by Chandra ACIS-I (red: 0.5-1~keV green: 1-2~keV blue: 2-8~keV).
Adaptively smoothing with a Gaussian kernal of $\sigma<10"$ has been applied. 21 X-ray point-like sources are detected in
this field. The properties of these sources are summarized in Table~1. 
The dashed ellipse illustrates the extraction region
for the remnant spectrum in both XMM-Newton and Chandra data (see Sec.~4).
}
\label{fig1}
\end{figure*}

\begin{figure*}
\centerline{\psfig{figure=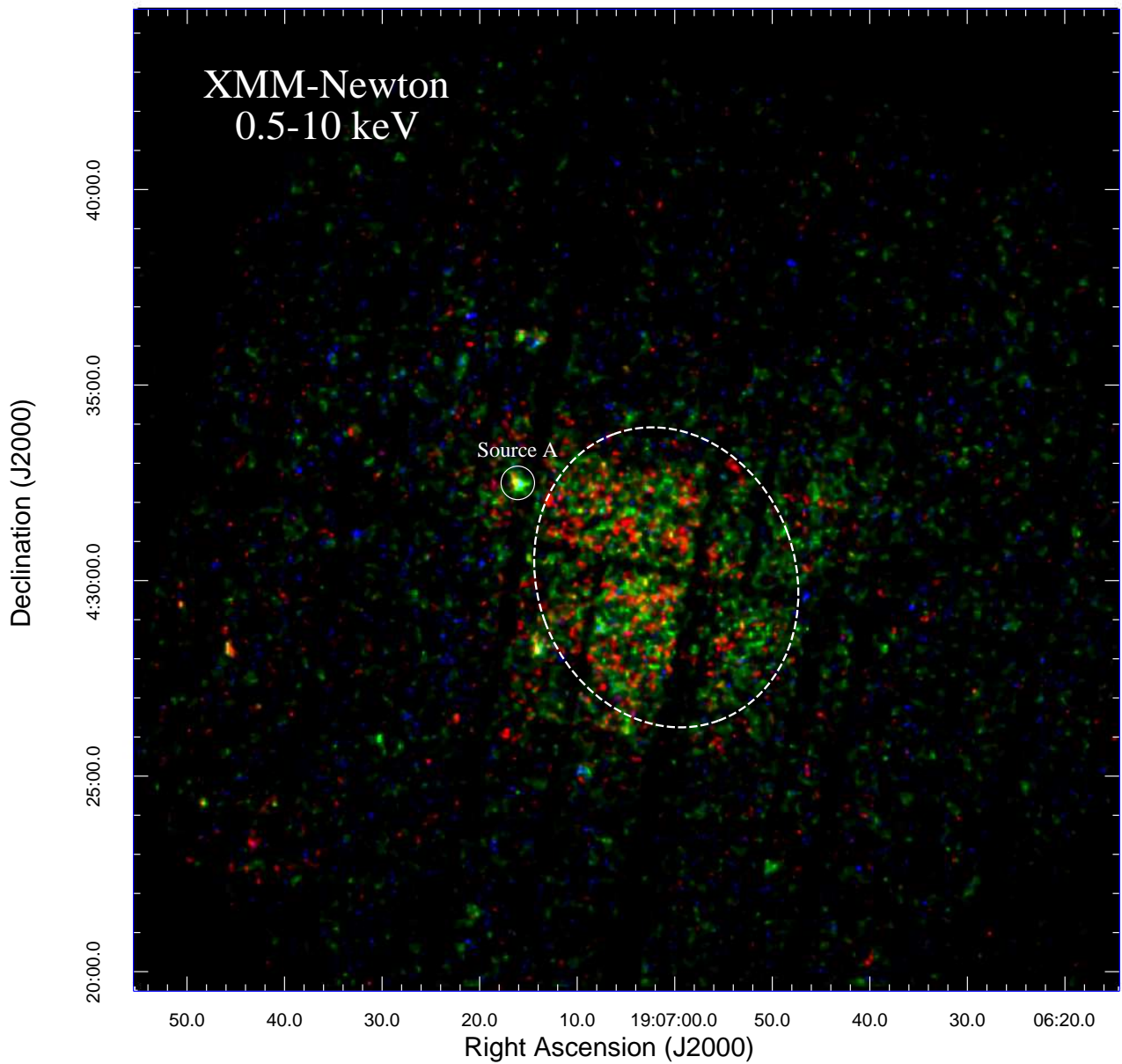,width=19cm,clip=}}
\caption{
Same field-of-view as Fig~\ref{fig1} as observed by XMM-Newton. The X-ray color
image (red: 0.5-1~keV green: 1-2~keV blue: 2-10~keV) is created by combining the data from all three cameras and
has been adaptively smoothed a Gaussian kernal of $\sigma<10"$. The dashed ellipse illustrates the extraction region
for the remnant spectrum in both XMM-Newton and Chandra data (see Sec.~4).
The only confirmed detection of point source in this XMM-Newton observation is
labelled as ``Source A" which is consistent with ``Source 20" detected in the Chandra field.
}
\label{fig1b}
\end{figure*}

Figure~\ref{fig1} and Figure~\ref{fig1b} display the X-ray color image of \snr\ as observed by Chandra 
in the energy range of 0.5--8~keV and XMM-Newton in the energy range of 0.5--10~keV respectively.  
These two images were created by using an adaptive smoothing algorithm with a Gaussian radius 
of $\leq 10''$ in order to probe the detailed structure of the diffuse emission. 
A center-filled X-ray morphology has been revealed. 
The angular extent of \snr\ as seen in X-ray is $\sim 8'\times 6.6'$
(major axis and minor axis of the dashed ellipse illustrated in Fig.~\ref{fig1}).\\[-2ex]

\begin{figure*}[t]
\centerline{\psfig{figure=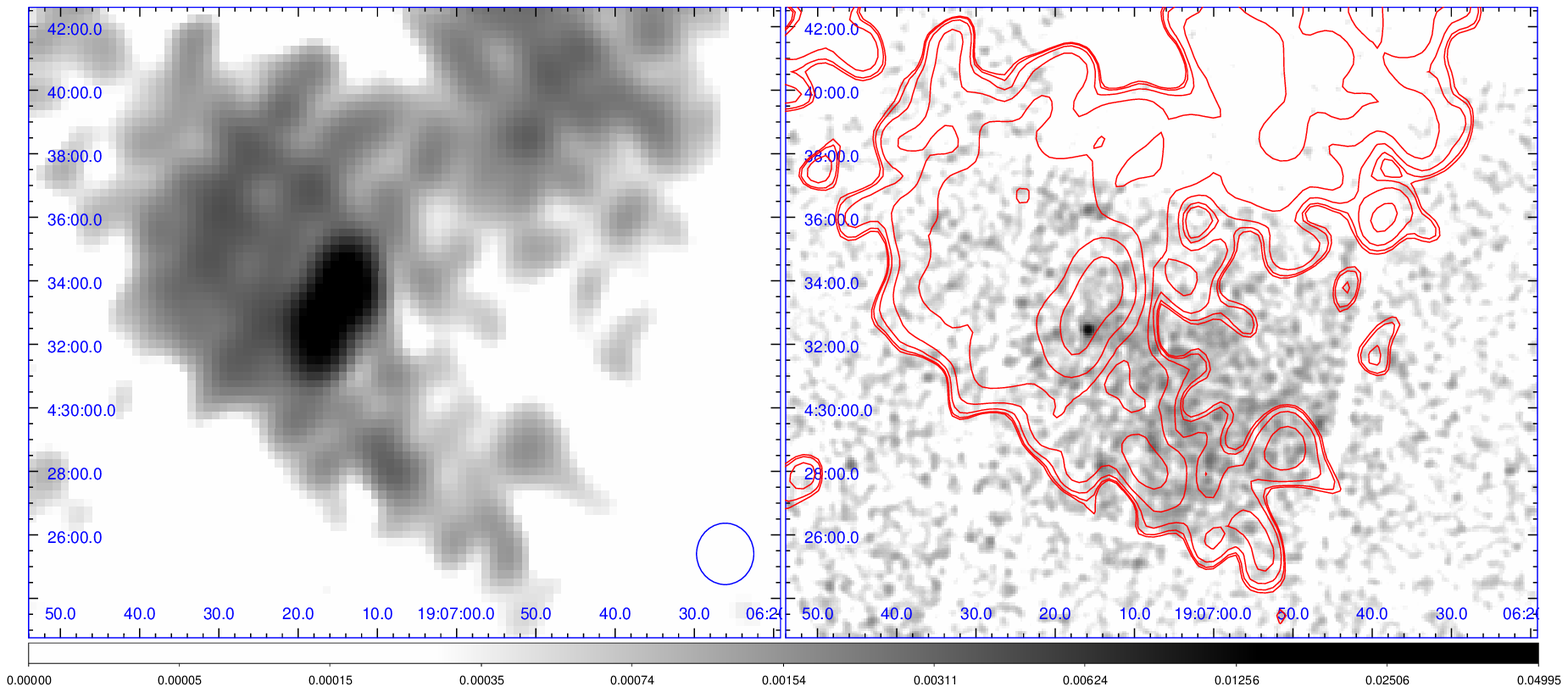,width=17cm,clip=}}
\caption{
{\em Left panel}: 1.4~GHz VLA image of \snr. The ellipse in the lower right 
corner indicates the restored beam size of $58'' \times 54''$. 
{\em Right panel:} Smoothed XMM-Newton MOS1/2 image of \snr\ superimposed with the 
radio contour determined as in the left panel. 
}
\label{fig3}
\end{figure*}

For investigating the radio counterpart of \snr\, we extracted the radio data taken with the 
Very Large Array (VLA) on August 18, 1996 from the NRAO Science Data Archive. 
These continuum observations were collected at a central frequency of 1.4~GHz in the 
VLA's D configuration. After standard cleaning process, a calibrated radio image of \snr\ 
is shown in Figure~\ref{fig3} left panel. This incomplete radio shell along the southern part of 
\snr\ is consistent with that taken with the Effelsberg radio telescope at 11~cm (Schaudel et al. 2002).
In this VLA radio map, apart from the diffuse emission, we noted an extended bright radio feature 
around the center of the FoV. For comparing the radio and X-ray features of \snr, we overlaid the 
radio contours on the XMM-Newton MOS1/2 image (see Figure~\ref{fig3} right panel). 

For searching the counterpart of \snr\ in the other wavelengths,
we have also explored the infrared data obtained by Wide-field Infrared Survey Explorer 
(WISE; Wright et al. 2010) and H$\alpha$ image downloaded from the Southern H-alpha Sky Survey Atlas 
(SHASSA; Gaustad et al. 2001). No possible diffuse infrared and H$\alpha$ emission associated 
with \snr\ was found in our study. \\[-2ex]

The sub-arcsecond resolution of Chandra enables us to search for the possible stellar remnant 
associated with \snr. By means of the wavelet source detection algorithm (CIAO tool: {\it wavdetect}), 
we searched for the point sources in the whole ACIS-I data. The exposure variation across the detector was 
accounted by the exposure map. 
We set the detection threshold such that no more than one false detection caused by background 
fluctuation is in the whole field. The lower limit of source significance was set to be 4$\sigma_{G}$, 
where $\sigma_{G}$ is Gehrels error. 21 sources were detected and marked by white circles in 
Figure~\ref{fig1}. The properties of these sources are given in Table~\ref{cha_src}. 
Based on the ratio between the source extents and the estimates of the PSF sizes at their position 
reported by {\it wavdetect}, most of these sources are found to be point-like except for Sources 13, 17 and 20. 
However, in view of their coincidence with the diffuse remnant emission, such ratios can possibly be overestimated. \\[-2ex]

To investigate if these X-ray sources are promising isolated neutron star candidates, we focused on 
those coinciding with the remnant emission (Source IDs: 13, 17, 19, 20) 
and proceeded to search for their possible optical counterparts since the X-ray-to-optical flux 
ratio ($f_{x}/f_{opt}$) provides a rudimentary parameter for discriminating the nature of the source. 
For an isolated neutron star, $f_{x}/f_{opt}$ is typically larger than 1000 (cf. Haberl 2007) 
while for field stars and active galactic nuclei the ratio are much lower than the typical $<0.3$ and 
$<50$, respectively (Maccacaro et al. 1988; Stocke et al. 1991).  
By cross-correlating the X-ray sources with the SIMBAD and NED databases and the VizieR 
catalogue service, we searched for the possible optical counterparts within a search radius of 
2 arcsec around each source. Among all these four sources, only Source 19 has an optical 
counterpart (USNO-B1.0 0944-0406013) identified. For the other three sources without optical counterparts 
found in this search, we computed their limiting X-ray to optical flux
ratios. The X-ray to optical flux ratio is defined as log($f_{x}/f_{opt}$)=log$f_{x}$+5.67+0.4R 
(Green et al. 2004), where R is the R-band magnitude and $f_{x}$ is derived in the 0.5--2.0~keV.
Since these X-ray sources are too faint for spectral fitting, we crudely estimated their X-ray fluxes 
in the 0.5--2.0~keV energy range from the count rates with the aid of WebPIMMS by assuming a 
weighted average column density of N$_{H}=1.0\times10^{22}$~cm$^{-2}$ toward \snr\ from 
the HI survey by Kalberla et al. (2005) and an absorbed power-law spectrum with $\Gamma=1.8$. 
Taking the limiting magnitude of the USNO-B1.0 catalog (i.e. R$>$21), the limiting X-ray to optical flux
ratios of Sources 13, 17 and 20 are found to be $>0.25$, $>0.26$ and $>0.32$ respectively. These values are not 
constraining in determining their emission nature. Dedicated optical observations of these sources are encouraged 
for further investigation. 

We have also computed their hardness ratios 
defined as HR=$(C_{2.5-8.0keV}-C_{0.5-2.5keV})/(C_{2.5-8.0keV}+C_{0.5-2.5keV}$), where 
$C_{0.5-2.5keV}$ and $C_{2.5-8.0keV}$ are the net counts in the 0.5--2.5~keV and 2.5--8.0~keV 
energy bands. The hardness ratio of Source 13, 17 and 20 are -0.46, 0.03 and -0.82 respectively, which are 
rather soft. If any of these sources are indeed isolated neutron stars, their emission should be
thermal dominant. 
\\[-2ex]

We have also attempted to search for the point sources with XMM-Newton data. A corresponding 
set of exposure maps was generated to account for spatial quantum efficiency, mirror vignetting, 
and field of view of each instrument by running the task {\em eexpmap}. Utilizing the SAS task 
{\em edetect\_chain}, we performed the source detection on MOS1, MOS2 and PN images individually. 
We set the threshold of detection likelihood to be {\em mlmin}=10 throughout the search, which 
corresponds to a detection significance of $\gtrsim$ 4-$\sigma$. Only one source, Source~A (see Fig.~\ref{fig1b}),  
can be detected by all three cameras.  
Its position consistent with Source 20 found by Chandra (see Fig.~\ref{fig1}). 
The inferior performance of source detection with XMM-Newton can 
be ascribed to its relatively poor resolution and high instrumental background.  
We also noted that the brightest part of the radio emission as seen by VLA conicides with Source A 
(see Figure~\ref{fig3} right panel). 

\begin{center}
\begin{deluxetable}{lcccccccc}
\tablewidth{0pc}
\tablecaption{X-ray sources in the field of view of Chandra ACIS} 
\startdata
\\ [-2ex]
\hline\hline
ID & RA (J2000) & Dec (J2000) & $\delta$RA\tablenotemark{a} & $\delta$Dec\tablenotemark{a}  & S/N\tablenotemark{b}  & Net counts &  Photon flux\tablenotemark{c}  & PSF Ratio\tablenotemark{d}  \\
           &  (h:m:s) & (d:m:s) &  (arcsec)  &   (arcsec)   &   $\sigma_{G}$   &  (counts) &  (10$^{-6}$~photon~cm$^{-2}$~s$^{-1}$)   & \\ 
\hline \\[-2ex]
  1 &  19:07:10.090 & +04:40:07.45   & 1.01  & 0.89  & 4.27  & $21\pm6$ &  $1.56\pm0.44$   & 1.74   \\
  2 &  19:07:35.220 & +04:35:42.74   & 0.76  & 0.75  & 4.33  & $23\pm6$ &  $1.81\pm0.50$   & 1.20   \\
  3 &  19:06:31.099 & +04:34:20.07   & 0.72  & 0.63  & 5.27  & $31\pm7$ &  $2.54\pm0.60$   & 0.95   \\ 
  4 &  19:06:46.800 & +04:29:32.99   & 0.14  & 0.28  & 6.68  & $40\pm8$ &  $3.50\pm0.71$   & 1.45   \\
  5 &  19:07:37.783 & +04:33:47.60   & 1.35  & 1.01  & 4.09  & $25\pm7$ &  $1.83\pm0.52$   & 2.59   \\
  6 &  19:06:40.222 & +04:27:06.79   & 0.89  & 1.05  & 4.01  & $22\pm6$ &  $1.73\pm0.51$   & 2.98   \\
  7 &  19:07:04.356 & +04:36:54.85   & 0.96  & 0.83  & 4.75  & $23\pm6$ &  $1.73\pm0.46$   & 3.49   \\
  8 &  19:07:35.777 & +04:31:00.48   & 0.54  & 1.13  & 5.03  & $26\pm6$ &  $1.96\pm0.49$   & 1.92   \\
  9 &  19:06:34.817 & +04:28:31.57   & 0.68  & 0.69  & 4.24  & $24\pm7$ &  $1.97\pm0.55$   & 1.70   \\
 10 &  19:06:43.714 & +04:27:53.03   & 0.42  & 0.36  & 4.27  & $22\pm6$ &  $1.59\pm0.45$   & 1.17   \\
 11 &  19:07:28.884 & +04:35:12.89   & 0.62  & 0.62  & 5.77  & $33\pm7$ &  $2.36\pm0.53$   & 1.83   \\
 12 &  19:07:32.844 & +04:31:12.61   & 0.65  & 0.76  & 4.97  & $26\pm7$ &  $2.03\pm0.51$   & 2.20   \\
 13 &  19:06:58.742 & +04:32:03.27   & 0.92  & 0.97  & 4.20  & $26\pm7$ &  $1.70\pm0.47$   & 28.21   \\
 14 &  19:06:48.403 & +04:38:06.99   & 0.86  & 0.77  & 6.59  & $36\pm7$ &  $2.58\pm0.53$   & 2.54   \\
 15 &  19:07:22.577 & +04:31:23.01   & 0.44  & 0.34  & 5.13  & $27\pm7$ &  $2.01\pm0.50$   & 1.92   \\
 16 &  19:07:15.862 & +04:36:14.94   & 0.62  & 0.58  & 6.22  & $28\pm6$ &  $1.87\pm0.42$   & 2.82   \\
 17 &  19:07:07.550 & +04:31:44.75   & 0.70  & 0.62  & 5.07  & $29\pm7$ &  $2.00\pm0.49$   & 19.69   \\
 18 &  19:07:14.371 & +04:36:03.77   & 0.36  & 0.34  & 7.65  & $43\pm8$ &  $3.14\pm0.58$   & 1.66    \\
 19 &  19:07:14.232 & +04:28:13.24   & 0.34  & 0.27  & 4.29  & $25\pm7$ &  $1.66\pm0.46$   & 2.14    \\
 20 &  19:07:16.094 & +04:32:30.02   & 0.55  & 0.59  & 6.69  & $34\pm7$ &  $2.24\pm0.46$   & 7.55   \\
 21 &  19:06:37.682 & +04:31:33.33   & 0.34  & 0.39  & 8.77  & $49\pm8$ &  $3.66\pm0.62$   & 1.68   \\
\enddata
\tablenotetext{a}{Position uncertainty.}
\tablenotetext{b}{Estimates of source significance in units of Gehrels error: 
$\sigma_{G}=1+\sqrt{C_{B}+0.75}$ where $C_{B}$ is the background counts.}
\tablenotetext{c}{Absorbed photon fluxes in the energy range of 0.5-8.0~keV.}
\tablenotetext{d}{The ratios between the source extents and the estimates of the PSF sizes.}
\label{cha_src}
\end{deluxetable}
\end{center}


\section{Spectral Analysis}

We extracted the spectrum of the remnant emission from the Chandra data within the elliptical region 
illustrated in Fig.~\ref{fig1}. All the detected point-like sources
within the selected region (i.e. Sources 4, 13 \& 17)
were removed before extraction. We utilized the CIAO tool {\em specextract} to extract 
the spectra and to compute the response files. In view of low signal-to-noise ratio, the intrinsic extent 
of \snr\ in X-ray is uncertain. Therefore, the background spectrum is sampled from the blank-sky events.
After background subtraction there are $\sim$2166 net counts available in 0.5-8~keV for the spectral analysis.
We binned the spectrum extracted from Chandra so as to have at least 20 counts per bin.

We utilized the XMM-Newton SAS tool {\em evselect} to extract the remnant spectrum from the XMM-Newton data within 
the same region adopted in the Chandra analysis. Although Sources 4, 13 and 17 were not 
detected by XMM-Newton, we have removed the photons within 15" from their positions in order to minimize the contamination. 
Similarly, the background spectra for each camera were sampled from the blank-sky events.
Response files were constructed by using the XMM-Newton SAS tasks {\em rmfgen} and {\em arfgen}. 
After background subtraction, there are 989, 1532 and 5990 net counts in 0.5-10~keV from MOS1, MOS2 and PN respectively. 
The extracted spectra were binned to have at least 15 for MOS1/2 and 30 for PN source counts per bin.

All the spectral fits were performed with the 
XSPEC software package (version: 12.7.0). All quoted errors are 1$\sigma$ for 2 parameters of interest.
For accounting the photoelectric absorption, we used the Wisconsin cross-sections throughout the analysis
(Morrison \& McCammon 1983; XSPEC model: WABS).\\[-2ex]

We began with examining the spectrum with an absorbed collisional ionization equilibrium (CIE) plasma model 
(XSPEC model: VEQUIL). First, we fixed the abundance of metals at the solar abundances 
in order to minimize the number of free parameters. We fitted the spectrum obtained from individual 
camera separately for checking whether the results are consistent. While the spectrum obtained 
from Chandra can be fitted reasonably well with the CIE model, residuals have been seen in all XMM-Newton 
spectra at energies $\gtrsim2$~keV. These residuals can be modeled by including an additional power-law model in the fitting
which yields a photon index of $\sim1.1$, $\sim0.3$ and $\sim0.7$ for MOS1, MOS2 and PN respectively. 
The steepness of these photon indices are inconsistent with being from a non-thermal component.
Therefore, we speculate that these residuals were possibly resulted from some residual soft proton 
contamination in individual cameras after the data 
screening. 

\begin{figure*}[t]
\begin{center}
\includegraphics[angle=-90,width=16cm]{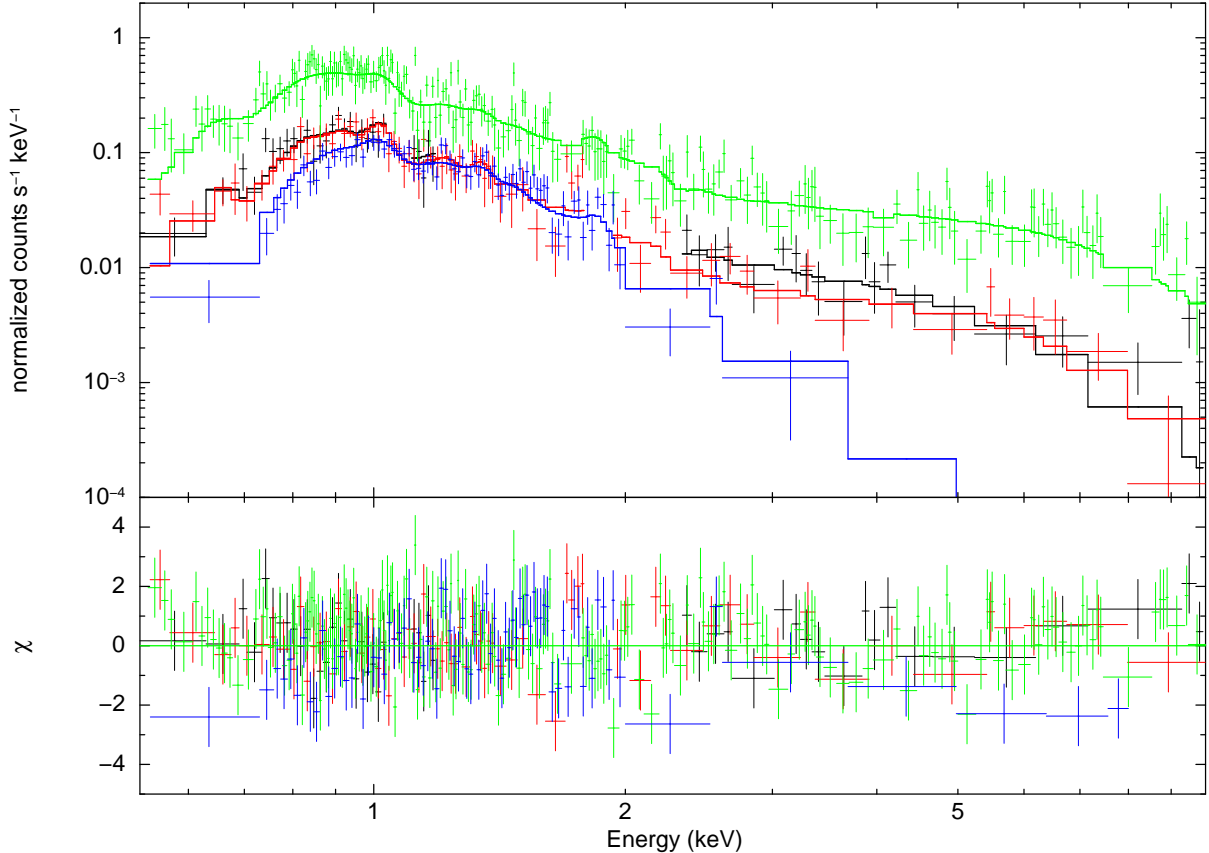}
\caption{  
({\em upper panel}) The X-ray energy spectra of \snr\ obtained from the XMM-Newton MOS1 (black), MOS2 (red), PN (green) 
and Chandra ACIS-I (blue) 
are simultaneously fitted with an absorbed collisional ionization equilibrium plasma model. Additional power-law component
have been applied to account for the residual soft proton contamination in the individual XMM-Newton camera.  
({\em lower panel}) Contributions to the $\chi^{2}$ fit statistic are shown.
}
\label{spec}
\end{center}
\end{figure*}

In order to tightly constrain the spectral parameters, we fitted the data obtained from all cameras simultaneously
with an untied power-law component to account for the residual soft proton contamination in the XMM-Newton data. 
The best-fit CIE model yields a column density of 
$N_{H}=6.85^{+0.33}_{-0.35}\times 10^{21}$~cm$^{-2}$ and a plasma temperature of $kT=0.64\pm0.02$~keV
(with $\chi^{2}$ = 842.47 for 704 degrees of freedom; hereafter dof). To examine whether 
the metal abundance of \snr\ deviates from the solar values, we thawed the corresponding parameters individually 
to see if the goodness-of-fit can be improved. With the abundances of oxygen (O) and neon (Ne) as 
free parameters, the fit is found to be somewhat improved ($\chi^{2}$ = 762.43 for 702 dof). Both O and 
Ne are suggested to be overabundant with respect to their solar values. The spectral parameters 
are tabulated in Table~\ref{spec_par}. The observed spectra with the CIE fit are display in Figure~\ref{spec}. \\[-2ex]

We also examined these spectra with an absorbed non-equilibrium ionization model of a constant 
temperature and a single ionization timescale (XSPEC model: VNEI) with the abundances of O and Ne 
as free parameters. The best-fit plasma temperature and the line-of-sight absorption are found 
to be $kT=0.65^{+0.03}_{-0.01}$~keV and $N_\mathrm{H}=5.44^{+0.33}_{-0.34}\times 10^{21}$~cm$^{-2}$.
The abundances of O and Ne are found to be $7.63^{+1.41}_{-1.30}$ and $3.39^{+0.84}_{-0.77}$ of their solar values. 
All these parameters are similar to those inferred from the CIE model. 
The inferred ionization timescale $\tau_{ion}$ 
is found to be $\sim6.8\times10^{12}$~s~cm$^{-3}$, which suggests the system can possibly reach the condition 
for CIE already. Also, statistically, the NEI model 
does not provide a better description of the data than the corresponding CIE fit ($\chi^{2}$ = 815.44 
for 701 dof). Therefore, we will not consider the NEI model in all subsequent discussions. \\[-2ex]

As the line features are not prominent in the observed energy spectra (see Fig.~\ref{spec}), we have also 
attempted to fit the data with a simple absorbed power-law model. Nevertheless, this fit yields an 
undesirable goodness-of-fit ($\chi^{2}$ = 1507.73 for 710 dof) and an unreasonable photon index of 
$\Gamma\sim4.8$. And hence, this emission scenario will not be further considered. 

In addition to the diffuse X-ray emission, we also examined the spectrum for the only point source 
detected by XMM-Newton, namely source A (see Fig.~\ref{fig1b}). The source spectrum has been extracted from a circular 
region with a radius of 15$''$ centered at the position reported by the source detection algorithm.  
This choice of extraction regions corresponds to the encircled energy fraction of $\sim$70\%. 
For the background subtraction, we sampled from the nearby low-count regions in the individual camera. 
There are $\sim$76 net counts in total available for the spectral fitting.\\[-2ex]

In view of its softness (see Sec. 3), we first examined the spectrum of Source A wtih a 
blackbody model (XSPEC model: BBODYRAD) which yields $N_\mathrm{H}=8.20^{+3.19}_{-2.17}\times 10^{21}$~cm$^{-2}$ 
and $kT=0.30^{+0.10}_{-0.07}$~keV
with an acceptable goodness of fit $\chi^{2}$ = 11.37 for 10 dof. The normalization implies an emitting
region with a radius of $R< 151.45D_\mathrm{kpc}$~m, where $D_\mathrm{kpc}$ is the source distance
in units of kpc. The unabsorbed flux of source A is found to be $f_{x} \simeq 3.9 \times 10^{-14}$~\fcgs\
in the energy range of 0.5-10.0~keV. We have also attempted to fit the spectrum with a power-law which results 
in similar a goodness of fit as the blackbody fit ($\chi^{2}$ = 11.89 for 10 dof). But we noted that the best-fit photon index 
($\Gamma=6.03^{+1.77}_{-1.33}$) is apparently steeper than a conventional range ($\Gamma\lesssim3$) of acceleration processes. 
We have also tested a scenario that it is a clump of the diffuse emission of \snr\ by fitting a VEQUIL model to its spectrum. 
However, the goodness of fit is worse than the other two tested models ($\chi^{2}$ = 12.29 for 10 dof).

\begin{table}[t]
\caption{X-ray spectral properties of \snr. All quoted errors are 1$\sigma$ for 2 parameters of interest.} 
\label{spec_par}
\begin{center}
\begin{tabular}{lc}
\hline\hline\\
\hline \\[-2ex]
$N_\mathrm{H}$ ($10^{21}$~cm$^{-2}$)     & $5.32^{+0.35}_{-0.32}$   \\[+1ex]
$kT$ (keV)            &  $0.65 \pm 0.03$         \\[+1ex]
O$^{a}$       &  $7.42^{+1.50}_{-1.46}$  \\[+1ex]
Ne$^{a}$     &  $3.49^{+0.86}_{-0.80}$  \\[+1ex]
Norm ($10^{-4})^{b}$  &  $6.32^{+0.95}_{-0.79}$  \\[+1ex]
$\chi^{2}$  & 762.43  \\[+1ex]
dof          & 702      \\
\tableline
\end{tabular}
\end{center}
$^{a}$ The metal abundance with respect to the solar value.\\ 
$^{b}$ The model normalization is expressed as $(10^{-14}/4 \pi D^{2}_{A})\int n_{e} n_{H} dV$, 
where $D_{A}$ is the angular diameter distance to the source (cm), $n_\mathrm{e}$ is the electron 
density (cm$^{-3}$), and $n_\mathrm{H}$ is the hydrogen density (cm$^{-3}$).\\
\end{table}

\section{$\gamma$-ray Observation \& Data Analysis}
In order to further probe if \snr\ is a site for GCR acceleration, we searched for the $\gamma$-ray 
emission at its location.
The Large Area Telescope (LAT) on board the Fermi Gamma-ray Space Telescope is able to detect 
$\gamma$-rays with energies between $\sim$100~MeV and $\sim$300~GeV (Atwood et al. 2009). 
Data used in this work were obtained between 2008 August 4 and 2012 November 21, which are available 
from the Fermi Science Support Center\footnote{http://fermi.gsfc.nasa.gov/ssc/data/analysis/software/}.
We used the Fermi Science Tools ``v9r23p1'' package to reduce and analyse the data in the vicinity of \snr. \\[-2ex]

Throughout this paper we used Pass 7 data with events in the ``Source'' class (i.e., event class 2) only. 
The corresponding instrument response functions (IRFs) ``P7SOURCE\_V6" (Atwood et al. 2009), which 
has been recommended for most analysis, was adopted. Photon energies were restricted to 200~MeV to 
300~GeV. To reject atmospheric $\gamma$-rays from the Earths limb, we excluded the events with 
zenith angles larger than 100$\degr$. In order to reduce systematic uncertainties and achieve a better 
background modelling, a circular region-of-interest (ROI) with a diameter of $10\degr$ centered at 
the nominal position of \snr\ was selected in our analysis. \\[-2ex]

To investigate the $\gamma$-ray spectral characteristic of \snr, we performed an unbinned likelihood 
analysis with the aid of \textit{gtlike}, by putting a point source with a PL model at the nominal position 
of \snr\ (i.e RA=$19^{\rm h}07^{\rm m}05^{\rm s}$ Dec=$+04^{\circ}31^{'}11^{''}$ (J2000)) . 
For the background model, we included the Galactic diffuse model  ({\tt gal\_2yearp7v6\_v0.fits}), 
the isotropic background ({\tt iso\_p7v6source.txt}), as well as all point sources reported in 
the Fermi LAT 2-Year Source Catalog (2FGL) within $10\degr$ from the center of the ROI. All these 
2FGL sources were assumed to be point sources which have specific spectra suggested by the 2FGL 
catalog (Nolan et al. 2012). While the spectral parameters of the 2FGL sources locate within the ROI 
were set to be free, we kept the parameters for those lying outside our adopted ROI fixed at the values 
given in 2FGL (Nolan et al. 2012). We allowed the normalizations of diffuse background components to 
be free. We found that there is no $\gamma$-ray detection of \snr\ in our study (TS=$-3\times10^{-3}$ at the nominal position 
of \snr). With the statistical
uncertainties of both photon index and the prefactor concerned, the $1\sigma$ limiting photon flux 
at energies $>100$~MeV is constrained to be $<1.3\times10^{-9}$~photons~cm$^{-2}$~s$^{-1}$.


\section{Discussion} 
Based on the best-fit X-ray spectral parameters, we discuss the physical properties of \snr. 
Our analysis suggests a plasma temperature of $\sim7.5\times10^{6}$~K which allows us to 
estimate the shock velocity $v^{2}_{s}=16kT/(3m_{p} \mu)$ (Reynolds 2008), where $k$ is the
Boltzmann constant, $m_{p}$ is the proton mass, $\mu$ is the mean mass per particle.
For a fully ionized plasma of cosmic abundances ($\mu\sim$0.6), 
the shock velocity is estimated to be $\sim745$~km~s$^{-1}$. \\[-2ex]

Assuming the shocked densities of hydrogen $n_\mathrm{H}$ and electrons $n_\mathrm{e}$ 
are uniform in the extraction region, the normalization of the CIE model can be approximated by 
$10^{-14} n_{e} n_\mathrm{H} V/4 \pi D^{2}$, where $D$ is the distance to \snr\ in cm and $V$ is 
the volume of interest in units of cm$^{3}$. Assuming a geometry of an ellipsoid for the extraction 
region, the volume of interest for the SNR is $\sim 1.29 \times 10^{56} D^{3}_\mathrm{kpc}$~cm$^{3}$, 
where $D_\mathrm{kpc}$ is the remnant distance in units of kpc. Assuming a completely ionized 
plasma with $\sim 10\%$ He ($n_{e} \sim 1.2 n_\mathrm{H}$), the 1-$\sigma$ confidence interval 
of the normalization implies that the shocked hydrogen and electron densities are in the ranges of 
$n_\mathrm{H} \simeq (0.21-0.24) D^{-0.5}_\mathrm{kpc}$~cm$^{-3}$ 
and $n_{e} \simeq (0.25-0.29) D^{-0.5}_\mathrm{kpc}$~cm$^{-3}$, respectively. \\[-2ex]

To determine the remnant age, we assume that \snr\ is in the Sedov phase and the shocked 
plasma is fully ionized with a single temperature. The shock temperature can be estimated by 
$T \simeq 8.1 \times 10^{6} E^{2/5}_{51} n^{−2/5}_\mathrm{ISM_{-1}} t^{−6/5}_{4}$~K (Hui et al. 2012), 
where $t_{4}$, $E_{51}$, and $n_\mathrm{ISM_{-1}}$ are the time after the explosion in units
of $10^{4}$~years, the released kinetic energy in units of $10^{51}$~erg, and the ISM density 
of 0.1 cm$^{-1}$, respectively. Assuming it is a strong shock, $n_\mathrm{ISM}$ 
is estimated to be 0.25$n_{H}$. Taking the 1-$\sigma$ uncertainties of the temperature inferred from 
the spectral fitting and assuming $E_{51}=1$ and a distance of 4~kpc, 
the age of \snr\ is constrained to be $(1.1 - 1.2) D^{1/6}_\mathrm{kpc} \times 10^{4}$~years. \\[-2ex]

Since the distance plays a crucial role in determining the physical properties of a SNR, to estimate
the distance of \snr\ is essential in our study. We first tried to obtain the distance at the lower side
via the optical extinction. Lucke (1978) built contour plots of equal mean reddening up to 2~kpc 
and found the the mean color excess $E_\mathrm{B-V}=0.25$~mag~kpc$^{-1}$ by using color 
excess and photometric distances in the UBV system for 4000 OB stars. From studies using ultraviolet 
spectroscopy of reddened stars and X-ray scattering halos in our Galaxy, Predehl \& Schmitt (1995) found the 
relationship between N$_\mathrm{H}$ and the total extinction A(V) to be approximately 
N$_\mathrm{H}$/A(V)=$1.79\times10^{21}$~cm$^{-2}$. Adopted the N$_\mathrm{H} \sim 5.3\times10^{21}$ 
inferred from the spectral fitting (cf. Table~\ref{spec_par}) and a typical value of 3.1 for A(V)/$E_\mathrm{B-V}$
in the Milky Way, we can crudely estimate a distance of $\sim$4~kpc. A dedicated HI observation of this newly 
identified SNR is encouraged for a more reliable distance estimation. \\[-2ex]

Non-detection of $\gamma-$ray emission in the energy range of $0.2-300$~GeV at the nominal position of 
\snr\ was reported in our study. 
We have placed a limiting photon flux of $F$($\ge$ 100 MeV)$<1.3\times10^{-9}$~photons~cm$^{-2}$~s$^{-1}$. 
Using the parameters inferred from the aforementioned X-ray analysis, we can estimate what is the expected intensity 
of the $\gamma-$ray flux from \snr. From Drury et al. (1994), the theoretical $\gamma-$ray flux can be estimated as 
$F$($\ge$ 100 MeV)$\approx4.4\times10^{-7}\theta\ (\frac{\rm E_{SN}}{10^{51}{\rm erg}})(\frac{d}{\rm 1~kpc})^{-2}(\frac{n}{\rm 1~cm^{-3}}) {\rm cm}^{-2}~{\rm s}^{-1}$. Adopting a distance of $d\sim4$~kpc, 
the ambient density of $n=n_{\rm ISM}\sim0.05$~cm$^{-3}$ as inferred from 
X-ray spectral fit and assuming a canonical explosion energy 
of $E_{\rm SN}\sim10^{51}$~ergs which converts $\theta=10\%$ into GCR energy (cf. Kang 2013), 
a flux of $F$($\ge$ 100 MeV)$\sim1.4\times10^{-10}$~${\rm cm}^{-2}~{\rm s}^{-1}$ is expected from \snr. 
A deeper $\gamma-$ray search 
is encouraged for further probing whether \snr\ is an acceleration site of GCRs indeed. 

\section{Summary \& Conclusion}

We have performed a detailed spectro-imaging X-ray study of the supernova remnant candidate 
\snr\ with XMM-Newton and Chandra. A central-filled X-ray structure correlated with an incomplete 
radio shell has been revealed. Its X-ray spectrum is thermal dominated and has shown the presence 
of a hot plasma accompanied with metallic emission lines. These observed properties indicate that 
\snr\ is a SNR belong to a mix-morphology category. 
The enhanced abundances of O and Ne suggest \snr\ might be resulted from a core-collapsed SN.
We have also searched for the possible 
$\gamma$-ray emission from \snr\ with \emph{Fermi} LAT data. With the adopted $\sim4.3$~yrs 
data span in this study, we report a non-detection of any $\gamma-$ray emission in the energy 
range of $0.2-300$~GeV.

\acknowledgments
This project is supported by the National Science Council of the Republic of China
(Taiwan) through grant NSC100-2628-M-007-002-MY3 and NSC100-2923-M-007-001-MY3.
CYH and KAS are supported by the National Research Foundation of Korea through grant 2011-0023383.
LT would like to thank the German \emph{Deut\-sche For\-schungs\-ge\-mein\-schaft (DFG)\/} 
for financial support in project SFB TR 7 Gravitational Wave Astronomy.



\end{document}